\documentclass[pre,aps,showpacs,amsmath,amssymb,twocolumn]{revtex4-1}

\usepackage{graphicx} 

\begin{document}

\title{Two rods confined by positive plates: Effective forces and charge distribution profiles} 

\author{G. Odriozola} 

\author{F. Jim\'{e}nez-\'{A}ngeles} 

\author{M. Lozada-Cassou} 

\affiliation{Programa de Ingenier\'{\i}a Molecular, Instituto Mexicano del Petr\'{o}leo, L\'{a}zaro C\'{a}rdenas 152, 07730 M\'{e}xico, D. F., M\'{e}xico}

\date{\today}
\begin{abstract}
Monte Carlo simulations were employed to study two negative rods confined between two parallel plates. The system is immersed in a 1-1 restricted primitive model electrolyte. Ion distributions and forces per unit of area (pressures) on rods and plates are accessed by sampling the $NVT$ ensemble. Pressures are analyzed by means of their corresponding electric and contact (depletion) contributions. This was done for several charge distributions on plates, plates charge densities, and plate-plate surface separation distances. We found an enhancement of the inherent repulsive rod-rod effective force when uncharged plates are the confining species. On the contrary, a strong decrease of the rod-rod effective repulsion was obtained for positively charged plates. Moreover, attraction was also found for plates having a charge equal to that of fully charged bilipid bilayers. These results agree with DNA-phospholipid complexation experiments. On the other hand, for a model having the plates charges fixed on a grid, very long range rod-rod sin-like effective forces were obtained. As explained in the text, they are a consequence of the rod-plate double layer coupling. 

\end{abstract}
\pacs{61.20.Qg,68.08.-p}

\maketitle

\section{Introduction}\label{intro}

In soft matter science, confinement is known to induce ordering, produce density fluctuations, provoke dielectric changes on liquids, affect diffusion processes, promote layering, enhance miscibility in polymer blends, and yield anomalous attractions between like charged colloidal particles \cite{Israelachvili,Levinger,Zhu,Odriozola_jcp2}. This last effect was observed experimentally by accessing the pair correlation function by means of video microscopy and optical tweezers \cite{Han}. Additionally, biological setups as those formed by plate-like charged lipid bilayers and rod-like DNA molecules show that, under certain conditions, confining bilayers induce a dense parallel 2D array of like-charged DNA molecules \cite{Wagner00,Radler97,Wu04M,Safinya01,Liang04}. Hence, it seems clear that confinement and self assembling are strongly linked.

The extensively used theory to explain interactions among charged colloidal particles is the one developed by Derjaguin, Landau, Verwey, and Overbeek (DLVO) \cite{Russel}. This is based on the Poisson-Boltzmann theory which deals with the electrostatic interaction among charged colloids and ions. This theory predicts that like charged colloids should repel each other following a screened Coulomb potential. There is, however, experimental evidence \cite{Kepler} and theoretical prove \cite{Lozada96} that under certain conditions two like charged macroparticles attract each other in bulk, and so, this mean field theory cannot predict this unintuitive behavior. On the other hand, there is also experimental evidence that like charged colloids, which do not attract in bulk, attract for the same conditions but under confinement \cite{Han}. Again, theories based on Poisson-Boltzmann cannot capture this peculiar behavior. Additionally, theories based on integral equations, which were success to predict the attraction of like charged colloids in bulk \cite{Marcelo96,Lozada97}, are of hard implementation for complicated geometries such as these produced by planar confinement. This is, at least partially, why there are to our knowledge not theoretical studies attempting to explain such anomalous behavior. 

Since geometry is not a problem for implementing simulations, this turns a natural tool for studying a system counting on confining charged plates, confined macroparticles, the corresponding counterions, and added salt. This is why we employ them to reveal the effect that confinement plays on like charged colloids effective interactions. Additionally, since lipid bilayers-DNA experiments show transitions from loose to compact DNA 2D ensembles by modulating the bilayers charge density, we would like to focus on modeling this particular system. For that purpose, two DNA molecules are modeled as two confined like charged negative rods, and the lipid bilayers as the confining positive charged plates. Hence, different plate-plate distances, charge on plates, and charge distributions on plates are studied, for a fixed 1-1 electrolyte concentration. 

The paper is structured as follows. Section \ref{intro} is this brief introduction. Section \ref{sim} gives details on how the simulations were performed. Results are presented in section \ref{res}, and section \ref{conc} summarizes the main contributions and tackles some conclusions.

\section{Simulations}\label{sim}

Monte Carlo (MC) simulations were performed to study a system counting on two infinitely large, parallel, negative rods confined by two parallel, positive plates of finite thickness. These hard macroparticles are immersed in a 1-1 restricted primitive model (RPM) electrolyte. The simulation box has sides lengths $L_x$$=$$L_z$$=$ 200 $\hbox{\AA}$, and $L_y$$=$ 125 $\hbox{\AA}$. The origin of coordinates is set at the box center. Plates are located parallel and symmetric to the $z$$=$$0$-plane, with a surface-surface separation distance $\tau$. Rods axes are placed parallel to the $y$ axis at $x$$=$$\pm (R+t/2),z$$=$$0$, being $R$ the rods radii and $t$ the rod-rod surface distance. 

For simplicity, plates, rods, and solution have the same dielectric constant. RPM consists of hard spheres with a centered point charge, such that their electrostatic interaction is 
\begin{equation} \label{elec}
U_{E}(r_{ij})=\frac{\ell _{B}z_i z_j}{\beta r_{ij}},
\end{equation}
being $\beta$$=$$1/k_BT$, $k_{B}$ the Boltzmann constant, $T$$=$$298$K the absolute temperature, $z_i$ and $z_j$ the valences of sites $i$ and $j$, $\ell _{B}$$=$$\frac{\beta e^2}{\epsilon}$$=$$7.14$$\hbox{\AA}$ the Bjerrum length, $\epsilon$$=$$78.5$ the dielectric constant, and $r_{ij}$ the interparticle distance. The electrostatic interaction among macroparticles and between macroparticles and ions are also given by equation \ref{elec} with $i$ and/or $j$ running over the macroparticles sites. 

We assigned to plates a charge density of $\sigma_0$$=$$0.229$$C/m^2$ ($+e$ per 70 $\hbox{\AA}^{-2}$) for each surface, in correspondence with a fully charged cationic lipid membrane \cite{Gelbart00}. To study the effect of varying the plates charge density, for some systems we assigned a fraction of this charge, expressed as a percentage. Three ways of distributing the charge on plates were explored. One consist of placing $+e$-charges in the center of 5 $\hbox{\AA}$ hard spheres, which are confined by the plates boundaries and are moved by following the MC criteria. The hardcore interaction of these 5 $\hbox{\AA}$ spheres is switched on just among themselves. The second way is by placing the $+e$-charges over triangular grids, which are randomly moved in the $x$ and $y$ directions to mimic a continuous charge distribution. These grids are placed on the surfaces of the plates. The third way is by placing the $+e$-charges over triangular grids, as in the previous case, but fixing them in these positions. On the other hand, rods have an $-e$-site each 1.7 $\hbox{\AA}$, as DNA molecules \cite{Gelbart00}. 

Rods have a radius of $R$$=$10.5 $\hbox{\AA}$, consistent with a hydrated DNA molecule. Plates are 5 $\hbox{\AA}$ thick to allow fluid-fluid correlations between the interplate and bulk sides of the plates \cite{Lozada96}. These correlations are expected in real 40 $\hbox{\AA}$ thick bilayers, due to their low dielectric constant. We fixed the concentration of the 1-1 electrolyte to $\rho_s$$=$0.1 M, and assigned its diameter to $a$$=$4.25 $\hbox{\AA}$, unless otherwise indicated. 

\begin{figure} \resizebox{0.4\textwidth}{!}{\includegraphics{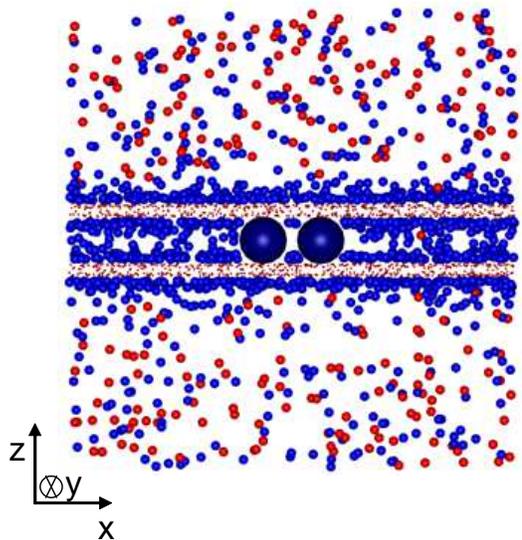}} \caption{\label{snapshot} Snapshot of an equilibrium configuration of a system counting on two rods sandwiched by two positive, confining plates. } \end{figure}

Additional anions are added to make the system electroneutral. Periodical boundary conditions are set for all directions. The Ewald summation formalism is employed to deal with Coulomb interactions \cite{Odriozola_lan,Alejandre94}. To access all phase space volume, {\it i.~e.}, to allow for ion interchange between the confined and unconfined regions, movements having a large maximum displacement are done. Figure \ref{snapshot} shows a front view snapshot of an equilibrated system having $\tau-2R$$=$0, $t$$=$a, and a 100$\%$ of plates charge. Here, the plates charge was distributed by the first method, blue means negative and red positive, and sizes are scaled except the 5 $\hbox{\AA}$ plate charges which are represented smaller.  

Electrostatic contributions to the forces acting on the macroparticles (rods and plates) are obtained by
\begin{equation}
\mathbf{F}_{el}=\langle \sum_i \sum_j -\nabla U_{E}(r_{ij}) \rangle,
\end{equation}
where $i$ runs over the sites of the reference macroparticle and $j$ runs over all other sites. The contact (depletion) contribution is obtained by integrating the ions contact density,$\rho(s$$=$cte$)$,
\begin{equation}
\mathbf{F}_c=-k_{B}T\int_s\rho(s=cte)\mathbf{n}ds,
\end{equation}
where $s$ refers to the macroparticle surface and $\mathbf{n}$ is a unit normal vector. These two contributions to the force are interdependent.

\section{Results}\label{res}

\begin{figure} \resizebox{0.4\textwidth}{!}{\includegraphics{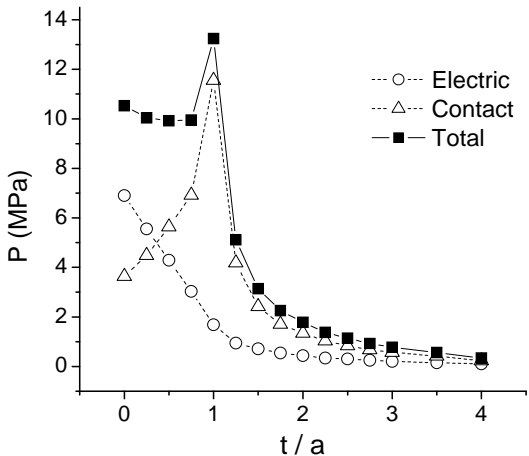}} \caption{\label{f-sinplan} Forces on rods per unit of rod area (pressure) as a function of $t$ for two unconfined rods.} \end{figure}

\begin{figure} \resizebox{0.48\textwidth}{!}{\includegraphics{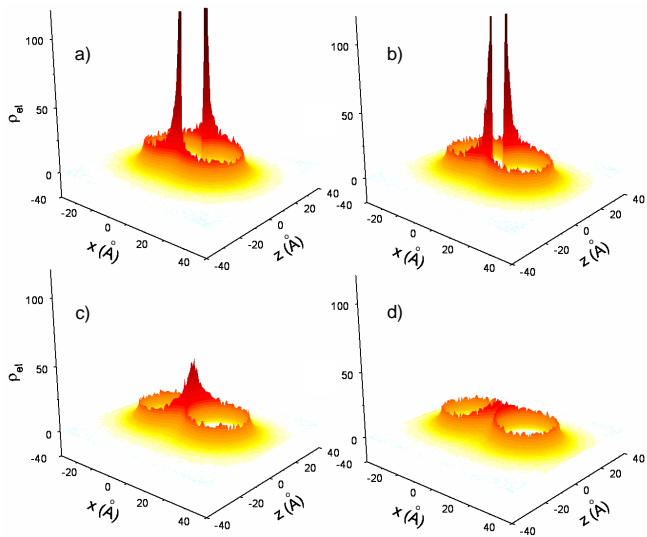}} \caption{\label{rho-sinplan} $\rho_{el}(x,z)$ as a function of $h$. Darker tones indicate larger absolute values of $\rho_{el}(x,z)$, while red means positive and blue negative.} \end{figure}

Before presenting the results obtained for confined rods, it is convenient to have in mind the most important features of the interaction between two parallel rods immersed in a 1-1 RPM electrolyte. This is why we show in figure \ref{f-sinplan} the effective forces per unit of rod area (pressure) as a function of the distance between rods surfaces, $t$, for two isolated rods immersed in a 1-1 electrolyte with a bulk concentration of 0.1 M. Note that system symmetry make the rod-rod forces to have only components on $x$. These components are shown in figure \ref{f-sinplan}, where a positive sign means repulsion. There, the total force results from the sum of two interdependent contributions: electric and contact (depletion). It is seen that both contributions are always repulsive and so it is the net force. The electric contribution is ever decreasing, since likely charged rods inherent repulsion decays with distance and since more counterions interpose between them as $t$ increase. When rods double layers turn independent the one another, both contributions become cero, as expected. The contact contribution, however, behaves differently. It increase with distance for $t<a$, and decays for $t<a$, showing a maximum for $t=a$. To understand this behavior, one should pay attention on how the charge distribution profiles around the macroparticles, $\rho_{el}(x,z)$, changes with $t$. This is defined by $\rho_{el}(x,z)$$=$$z_{+}g_{+}(x,z;\tau,t)+z_{-}g_{-}(x,z;\tau,t)$, being $g_{+}(x,z;\tau,t)$ ($g_{-}(x,z;\tau,t)$) the cation (anion) distribution profiles, given $\tau$ and $t$. This case corresponds to $\tau \rightarrow \infty$ (no plates in the system). Figure \ref{rho-sinplan} shows $\rho_{el}(x,z)$ for $t$$=$0, 0.75a, 1.5a, and 2.25a. For $t$$<$a (plots a) and b)) figure \ref{rho-sinplan} shows large $\rho_{el}$ values at $(x=0, z=\pm \sqrt{R_a^2-(R+t/2)^2})$, being $R_a$$=$$R+a/2$. This indicates the formation of two lines of counterions along the rods dumbbell. These counterions pressure on both rods producing strong repulsive contact contributions, which are not compensated by the external counterions \cite{Felipe4}. Hence, as $t$ increases, the lines of counterions approach each other producing larger contact contributions on the $x$ axis. This explains the increasing trend of the contact contribution with $t$ for $t<a$. For $t>a$, the peaks at $(x=\pm t/2, y=0)$ decreases with $t$ (figures \ref{rho-sinplan} c) and d)) and so it does the contact contribution. This produces the maximum of the contact component at $t=a$, as well as the maximum of the total pressure. Finally, it should be mentioned that the work per unit of charged site necessary to move the rods from $t$$=$4a to $t$$=$0 is 2.13 $K_BT/site$.  

\subsection{Influence of confinement}

Keeping all the same conditions fixed but including the plates in the system, we studied how confinement influences the rod-rod effective interaction. As a first experiment, we considered the plates to be fully charged (100$\%$), having the charge distributed according to the first method (charge sites are confined by plates boundaries but are moved by following the MC criteria). This model may describe the distribution of phospholipid charged heads inside the bilayer (plates). 

\begin{figure} \resizebox{0.4\textwidth}{!}{\includegraphics{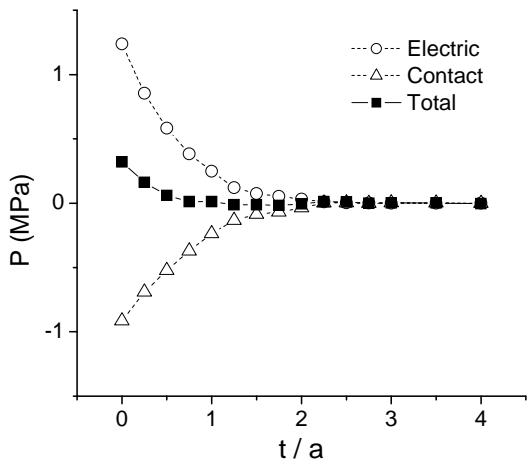}} \caption{\label{f-cabmov} Forces on rods per unit of rod area (pressure) as a function of $h$ when highly charged confining plates are present. } \end{figure}

\begin{figure} \resizebox{0.48\textwidth}{!}{\includegraphics{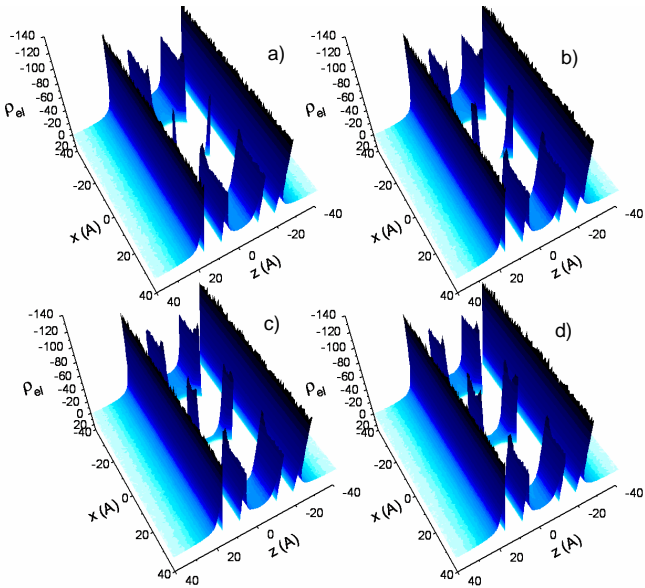}} \caption{\label{rho-cabmov} $\rho_{el}(x,z)$ for $t$$=$0 and $h$$=$ 0, 0.75a, 1.5a, and 2.25a. Darker tones indicate larger absolute values of $\rho_{el}(x,z)$, while red means positive and blue negative. } \end{figure}

The forces obtained by means of this experiment are shown in figure \ref{f-cabmov}. They strongly contrast with the previously shown in many aspects. To begin with, all contributions are very close to cero for all distances, $t$. In particular, the total force is practically negligible due to the counterbalance of a small repulsive electric contribution and a small attractive contact contribution. This leads to a work per unit of charged site necessary to move the rods from $t$$=$4a to $t$$=$0 of 0.01 $K_BT/site$. In addition, it seems to be attractive in the region $1.25a$$\leq$$t$$\leq$$2a$, and clearly positive for $t$$\leq$$a$. In fact, we obtain a very small but negative work per unit of rod site necessary to move the rods from $t$$=$4a to $t$$=$1.25a, {\it i.~e.}, -0.0005 $K_BT/site$. Hence, strong changes in the charge distribution profiles around the macroparticles are expected. They are shown in figure \ref{rho-cabmov} for $h$$=$0, 0.75a, 1.5a, and 2.25a. As can be seen, the charge density surrounding the rods, which used to be positive, is now negative. It should be noted that this case corresponds to the one shown in figure \ref{snapshot}, where just a cation is seen in the confined region for this particular configuration. Hence, cations are expelled out from the sandwiched region, in favor of anions, which are the plates counterions. This large anion concentration in the interplate region counterbalances the rod-rod electrostatic repulsion, producing a remarkable decrease of the electrostatic contribution. On the other hand, the large cation peaks at $(x=0, z=\pm \sqrt{R_a^2-(R+t/2)^2})$, which were responsible for the large contact repulsive contribution, are not present any more, and so it is the contact repulsive behavior. Moreover, the anions peaks that grow outside the rods and close to the plates overcompensates the anion peaks at $(x=0, z=\pm \sqrt{R_a^2-(R+t/2)^2})$ producing an attractive contact contribution. These outside peaks are seen in figure \ref{rho-cabmov} for all distances $t$. It is also seen that the inner peaks at $(x=0, z=\pm \sqrt{R_a^2-(R+t/2)^2})$ are smaller for figures \ref{rho-cabmov} a) and b). In fact, figure \ref{rho-cabmov} a) shows very narrow peaks at $(x=0, z=\pm \sqrt{R_a^2-(R+t/2)^2})$ indicating that anions can barely enter in these inter rod regions. 

Figure \ref{rho-cabmov} also shows large interplate ionic depletion regions due to the presence of the rods. As expected, these depletion regions unbalance the contact forces inside and outside the plates, producing an attractive net contact force per unit of plate area of -6.72 MPa. This force is practically independent of the rods positions (the informed value is the average). The electric contribution to the force on plates is repulsive but practically cero, {\it i.~e.}, 1.12 MPa. This leads to a total force per plate area of -5.6 MPa. 

\begin{figure} \resizebox{0.4\textwidth}{!}{\includegraphics{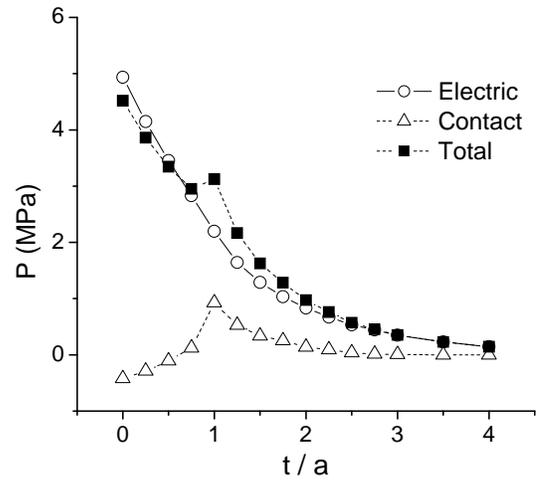}} \caption{\label{f-sep-a} Forces on rods per unit of rod area (pressure) as a function of $h$ for $\tau-2R$$=$$2a$. } \end{figure}

\begin{figure} \resizebox{0.48\textwidth}{!}{\includegraphics{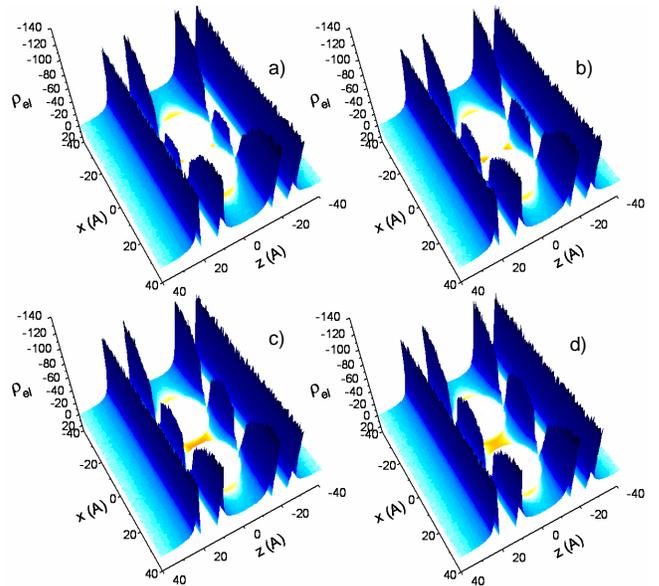}} \caption{\label{rho-sep-a} $\rho_{el}(x,z)$ for $t$$=$a and $h$$=$ 0 (a), 0.75a (b), 1.5a (c), and 2.25a (d). Darker tones indicate larger absolute values of $\rho_{el}(x,z)$, while red means positive and blue negative.} \end{figure}

A second computer experiment was carried out for the same conditions but $\tau-2R$$=$$2a$. Under these conditions, the forces per unit of rod area are plotted in figure \ref{f-sep-a}. These forces show an intermediate behavior between cases $\tau-2R$$=$$0$ and $\tau-2R$$\rightarrow$$\infty$ (compare figures \ref{f-sinplan} and \ref{f-cabmov} with figure \ref{f-sep-a}). That is, both contributions to the force were diminished by confinement, but in this case, the decrease is not as pronounced as for $\tau-2R$$=$$0$. Indeed, the work per unit of rod sites to move the rods from 4a to 0 is 0.78 $KbT/site$, which lies in between the values obtained for unconfined and confined conditions. Additionally, other features such as the contact force peak at $t$$=$$a$ persist, which is a signature of the presence of rods counterions in between the rods, as found for $\tau-2R$$\rightarrow$$\infty$. Nevertheless, the negative contact contribution for short rod-rod distances indicates the presence of large anions peaks outside the rods, as found for $\tau-2R$$=$$0$. 

These features are depicted in figure \ref{rho-sep-a} where the charge density profile is shown for $t$$=$$0$, 0.75a, 1.5a, and 2.25a. There, the yellow regions close to the rods and at the midplane indicate the presence of rods counterions (cations). It is observed that they enter in between the rods as far as their surface-surface distance increase. This is similar to case $\tau-2R$$\rightarrow$$\infty$. In addition, anion peaks outside the rods are larger than those in between them for $t$$=$0 and 0.75a (figures \ref{rho-sep-a} a) and b)). Another important difference with respect to case $\tau-2R$$=$$0$ is that interplate ionic depletion regions are smaller. This impacts on the contact plate-plate pressure, which in this case is -3.6 MPa. The electric contribution turns attractive, {\it i.~e.}, is -1.0 MPa, and the total pressure is -4.6 MPa, for $t=0$. In this case, the attractive pressure on plates decreases with $t$ reaching values close to -3 MPa. 

\subsection{Influence of charge distribution on plates}

\begin{figure} \resizebox{0.4\textwidth}{!}{\includegraphics{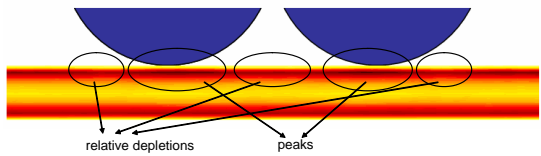}} \caption{\label{rho-plan} $\rho^*_{el}(x,z)$ for planes. Darker tones indicate larger absolute values of $\rho_{el}(x,z)$.} \end{figure}

It was shown in the previous section that confinement strongly affects the interaction between two macroparticles (rods, in our case). In the present section, we would like to elucidate the influence of the charge distribution on plates on the rod-rod effective force. To that purpose, we should first pay attention on the positive charge distribution on plates, $\rho^*_{el}(x,z)$, obtained for case $\tau-2R$$=$$0$, $t=0$. Here, $\rho^*_{el}(x,z)$ is defined by $z_{*+}g_{*+}(x,z;\tau,t)$, being $z_{*+}$ the valence and $g_{*+}(x,z;\tau,t)$ the distribution profiles of charged sites inside the plates, given $\tau$ and $t$. For these conditions, $\rho^*_{el}(x,z)$ is shown in figure \ref{rho-plan} for the lowermost plate, where the rods are schematized for clarity. 

It can be seen that, driven by energy and entropy, the plates charges move toward the plates surfaces forming almost homogeneously distributed charge surface densities, which are not very influenced by the rods positions. This is maybe the most important conclusion, since it makes us think that a homogeneous surface charge distribution may also model this system. On the other hand, it is also true that a small deviation from an homogeneously surface charge distribution is obtained. That is, charge peaks close to the rods are observed, which in turn produce relative depletion regions at their sides. These features are highlighted in figure \ref{rho-plan}. 

\begin{figure} \resizebox{0.4\textwidth}{!}{\includegraphics{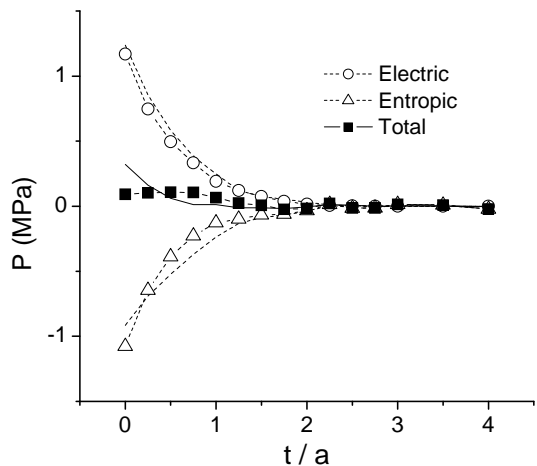}} \caption{\label{f-cabmov-movplan} Forces on rods per unit of rod area (pressure) as a function of $h$ for a continuously like surface charge distribution on plates. The lines corresponds to the results presented in figure \ref{f-cabmov}} \end{figure}

Hence, let us test a simpler model where the positive charge on plates is distributed on triangular grids at both plates interfaces, which are frequently, independently, and randomly moved in the $x$ and $y$ directions, to mimic homogeneous surface charge distributions. The forces on rods obtained by using this model are presented in figure \ref{f-cabmov-movplan} and compared with those obtained by moving the charged plate sites by following the MC criteria. As can be seen, in general, a very similar behavior is obtained. That is, both entropic contributions are negative, monotonously increasing functions of $t$, yielding cero for $t$$>$$2.5a$, and both electric contributions are positive, ever decreasing functions of $t$, producing cero values for $t$$>$$2.5a$. Additionally, both total contributions yields practically cero values for all $t$, being positive for $t$$<$$a$ and negative for $1.25a$$<$$t$$<$$2.25a$. The main difference seems to be the smaller value of the total force at rod-rod contact obtained for the continuously like surface charge distribution model. On the other hand, the obtained value of the total plate-plate force is -5.5 MPa, which compares well with that obtained for movable plates charged sites (-5.6 MPa). Hence, we may conclude that, at least for these conditions, the discrete nature of the bilayers charge should not strongly affect the forces on DNA molecules. 

\begin{figure} \resizebox{0.4\textwidth}{!}{\includegraphics{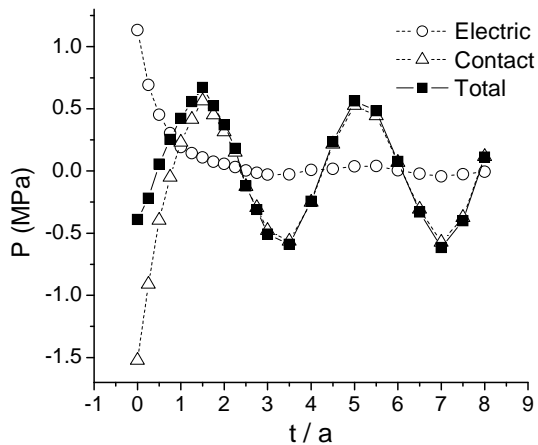}} \caption{\label{f-cfij} Forces on rods per unit of rod area (pressure) as a function of $h$ for a continuously like surface charge distribution on plates.} \end{figure}

Leaving aside the small differences between the previous studied models, we still would like to know whether a structured fixed grid of discrete surface charge affects the rod-rod effective interaction. This, of course, is not expected to be a good model for a lipid bilayer, but may be a model of certain type of nanostructured layers. Results are shown in figure \ref{f-cfij}. For very small $t$, they are similar to those shown by figure \ref{f-cabmov-movplan}, {\it i.~e.}, there is an attractive decaying contact contribution and a repulsive decaying electrostatic contribution. For $t$$>$ 0.5a, however, the contact contribution crosses over yielding positive values. In addition, it produces a maximum at $t$$=$1.25a. Moreover, it produces positive maxima and negative minima at every 7.8 $\hbox{\AA}$, showing a sin-like behavior. Note that these maxima and minima reach similar absolute values, independently of how large the rod-rod distance is. In other words, this behavior is very long ranged and so, it seems to be a consequence of the rod-plates interaction, indeed. On the other hand, the electric contribution shows also a sin-like behavior, once the rods are far enough the one another. Its amplitude is much smaller than the contact contribution, but seems to be in phase with it. 

\begin{figure} \resizebox{0.48\textwidth}{!}{\includegraphics{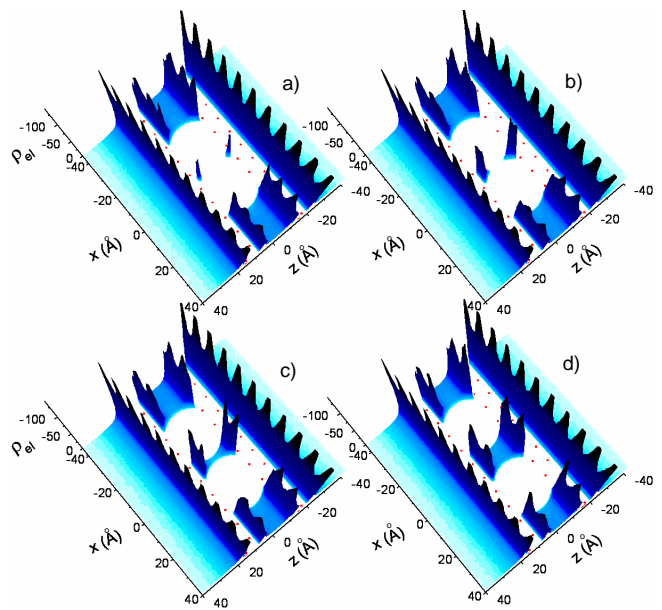}} \caption{\label{rho-cfij} $\rho_{el}(x,z)$ for a discrete surface charge distribution on plates, $\tau-2R$$=$0, and $h$$=$ 0 (a), 0.75a (b), 1.5a (c), and 3a (d). Darker tones indicate larger absolute values of $\rho_{el}(x,z)$. Red dots in plates indicate where the discrete positive charge is located. } \end{figure}

The rod-rod effective forces must be related to the charge distribution profiles of the interplate region. This is shown in figure \ref{rho-cfij} for $h$$=$0, 0.75a, 1.5a, and 3a. In all plots of figure \ref{rho-cfij} are seen peaks and valleys of the charge distribution close to plates. Naturally, peaks coincide with the locations of the positive charge of plates and valleys are between peaks. Hence, the peak to peak distance is equal to the $x$ distance between adjacent charged plates sites, $\lambda$$=$7.8 $\hbox{\AA}$. Note that the locations of the internal charged sites of both plates are the same by construction, and so, internal peaks and valleys of the charge distribution profile of one plate face those of the other. This produces a reinforcement of the effects that peaks and valleys produce on the rod effective forces. For instance, in figure \ref{rho-cfij} a) it is observed that the anionic distribution at contact with the internal (closer to the other rod) surface of rods corresponds to valleys and outside to peaks, explaining the attractive contact force at $t$$=$0. In figure \ref{rho-cfij} b), the inside and outside charge distribution at rod surfaces do not correspond to peaks or valleys, and hence, contact forces correspond neither to maximum nor to a minimum ($t$$=$0.75a). For $t$$=$1.5a a maximum is observed for the contact force, which is explained by the ionic peaks at contact with the internal surface of rods (see figure \ref{rho-cfij} c)), and the valleys at contact with the external rods surfaces. The opposite situation is found for $t$$=$3a, where peaks are found in contact with the outside of rods and valleys at the inside (see figure \ref{rho-cfij} d)). Hence, it turns clear that this apparent long range rod-rod interaction is nothing but a short range rod-plate double layer coupling effect. 

It must be noted that this kind of effects are a function of $\sqrt(R^2-(R-a/2)^2)/\lambda$ and of the relative position between the charged sites of one plate and the other. For instance, if the charged sites of one plate and the other had not been face-to-face but unphased a distance $\lambda/2$, none of these effects would have occurred. None of these effects would have happened for $\sqrt(R^2-(R-a/2)^2)/\lambda$$=$0.5 either. On the contrary, these effects are expected to be strong for $\sqrt(R^2-(R-a/2)^2)/\lambda$$=$1. Our system has $\sqrt(R^2-(R-a/2)^2)/\lambda$$=$ 0.82 and has the charged sites of one plate faced to the charged sites of the other, and this is why the effects are clearly seen. Another thing to note is that the $x$ position of one plate with respect to the other controls the phase of the two lines of peaks and valleys of the charge distribution profiles close to plates, which in turn controls the apparent rod-rod long range interaction. Hence, by changing the plates relative positions the rod-rod separation distance should change accordingly. This property may be of practical interest for nanostructure manipulation. 

\subsection{Influence of plates charge density}

\begin{figure} \resizebox{0.4\textwidth}{!}{\includegraphics{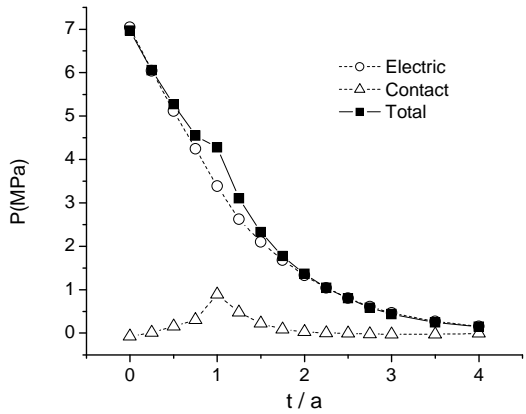}} \caption{\label{f-perc50} Forces on rods per unit of rod area (pressure) as a function of $h$ for plates having a 50 $\%$ charge density. } \end{figure}

DNA-lipid bilayer lamellar phase ensembles show a transition from condensed to loose DNA 2D arrays modulated by the rate of cationic (di-oleoly thrmethylammonium) to neutral (dioleoyl phosphatidylcholine) lipid molecules, which form part of the confining bilayers. Experiments show that for a fully charged membrane, {\it i.~e.}, no neutral lipid added, DNA form a condensed arrengement which has a DNA-DNA distance of 24.5 $\hbox{\AA}$. This distance enlarges to 57.1 $\hbox{\AA}$ for 57$\%$ of the full charge density. Hence, we find interesting to study the rod-rod effective interaction for smaller plates charge densities. 

Results for half the full charge density of plates are shown in figure \ref{f-perc50}. The total effective rod-rod force is a positive monotonically decreasing function of the rod surface distance, {\it i.~e.}, its always repulsive. Hence, the work per unit of rod site to bring them in touch is 1.15 $K_BT/site$. It should be noted that this value is relatively similar to that obtained for fully charged membranes but for a larger interplate distance (0.78 $K_BT/site$). In fact, all force components of figure \ref{f-perc50} are very similar to those shown in figure \ref{f-sep-a}. On the other hand, the pressure on plates is -5.1 MPa for the rods at contact, and decreases its absolute value to -4.5 MPa for larger rod-rod distances. In this case, the electric component is the larger contribution being -4.8 at $t$=0, and decreasing to yield -3.2 for $t$=4a. 

\begin{figure} \resizebox{0.48\textwidth}{!}{\includegraphics{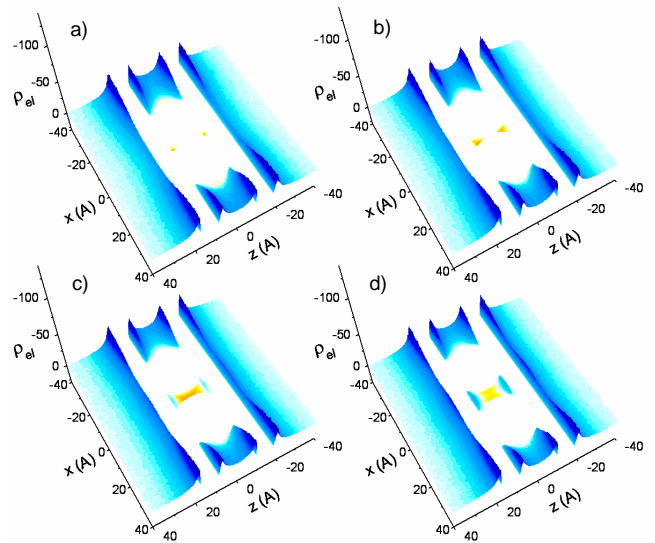}} \caption{\label{rho-perc50} $\rho_{el}(x,z)$ for 50 $\%$ charged plates, $\tau-2R$$=$0, and $h$$=$ 0 (a), 0.75a (b), 1.5a (c), and 2.25a (d). Darker tones indicate larger absolute values of $\rho_{el}(x,z)$. } \end{figure}

The corresponding charge densities are shown in figure \ref{rho-perc50}. As expected, much lower values of the charge density at contact with the plates surfaces are obtained, as a consequence of the lower charge of the plates. Hence, the interplate region becomes less dense in anions (plates counterions) and positive charge densities are observed close to the rods surfaces. In particular, cations are allowed to fill the inter rod regions at $(x=0, z=\pm \sqrt{R_a^2-(R+t/2)^2})$, which are responsible for the contact characteristic peak at $t=a$ (similar to figures \ref{f-sinplan} and \ref{f-sep-a}). Another consequence of a less crowded anionic interplate is the increase of the rod-rod effective repulsion. That is, the less counterions in the interplate, the less compensated the inherent rod-rod repulsion becomes. Finally, other point to note is that the charge densities at contact with the outer plates surfaces becomes clearly dependent on the $x$ coordinate. Since in this case the charge on plates do not overcompensate that on rods, the presence of the rods strongly affects the outer charge distribution. 

\begin{figure} \resizebox{0.4\textwidth}{!}{\includegraphics{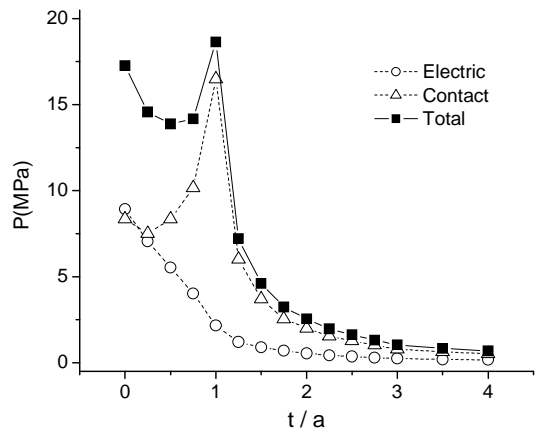}} \caption{\label{f-perc0} Forces on rods per unit of rod area (pressure) as a function of $h$ for uncharged plates. } \end{figure}

When confinement is produced by uncharged plates the rod-rod total effective force becomes very repulsive, mainly due to the increase of the contact contribution. This is clearly seen in figure \ref{f-perc0}, where the large peak of the contact contribution at $t=a$ makes the total pressure to reach values close to 20 MPa. The general trend of the contributions is, however, similar to that already shown for unconfined conditions. In addition, the obtained pressure for plates is 2.4 MPa for $t=0$, which decrease for increasing $t$ yielding 0.4 MPa for $t=4a$. Obviously, the total pressure coincides with the contact contribution, since a cero electric contribution is obtained. Due to the fact of obtaining large positive plate-plate pressures, a configuration where rods are squeezed by plates is, in this case, not expected. 

\begin{figure} \resizebox{0.48\textwidth}{!}{\includegraphics{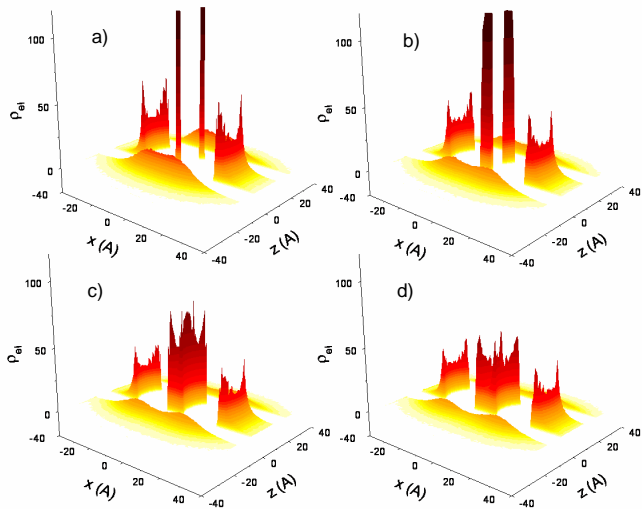}} \caption{\label{rho-perc0} $\rho_{el}(x,z)$ for uncharged plates, $\tau-2R$$=$0, and $h$$=$ 0 (a), 0.75a (b), 1.5a (c), and 2.25a (d). Darker tones indicate larger absolute values of $\rho_{el}(x,z)$. } \end{figure}

Confinement by uncharged plates also affects the charge distribution profiles. The fact that no ions can attach to part of the rods surfaces due to the plates excluded volume, makes the system react by pushing more ions to the available surfaces. In particular, those pore-like regions (large surface/volume rate), as those located at $(x=0$,$z=$ $\pm$ $\sqrt{R_a^2-(R+t/2)^2})$, turn crowded by cations (rods counterions). Again, these cations exert large repulsive contact contribution on rods, explaining the huge effective force peak at $t=a$.

\section{Conclusions}\label{conc}

Monte Carlo simulations were performed to study a system counting on two charged, parallel rods, and two confining charged, parallel plates. The system is immersed in a 1-1 restricted primitive model electrolyte. Effective forces acting on the macroparticles and the corresponding charge density profiles were accessed for different confinement conditions, {\it i.~e.}, the influence of the amount of charge of the confining plates, its distribution, and the plate-rod distance were studied). Both situations, an enhancement of the inherent repulsive rod-rod effective force, and a strong decrease were obtained for uncharged and charged plates, respectively. Some cases even show attraction, as found for highly charged confining plates. Hence, highly charged plates promote rod-rod crowding, and lower charged plates induce larger rod-rod separation distances. This well agrees with DNA-bilipid experiments. We found that models having movable charged sites on plates and homogeneously-like charged surfaces lead to very similar results. 

On the other hand, a model having discrete, fixed site charges on plates, located forming a regular grid, induce a very long range rod-rod effective interaction. This long range interaction shows a sine-like behavior, having several maxima and minima regularly separated. As pointed out in the paper, it is a consequence of the double layer coupling between rods and plates, and has little to do with the rod-rod direct correlation. This is why we prefer to qualify it as apparent long range rod-rod interaction. Since this interaction rules the rod-rod separation distance and since the interaction depends on the relative position between plates, controlling this relative position, in turn, one controls the degree of compactness of the parallel stack of rods. This may be of practical use for nanoconstruction purposes.

\begin{figure*} \resizebox{0.85\textwidth}{!}{\includegraphics{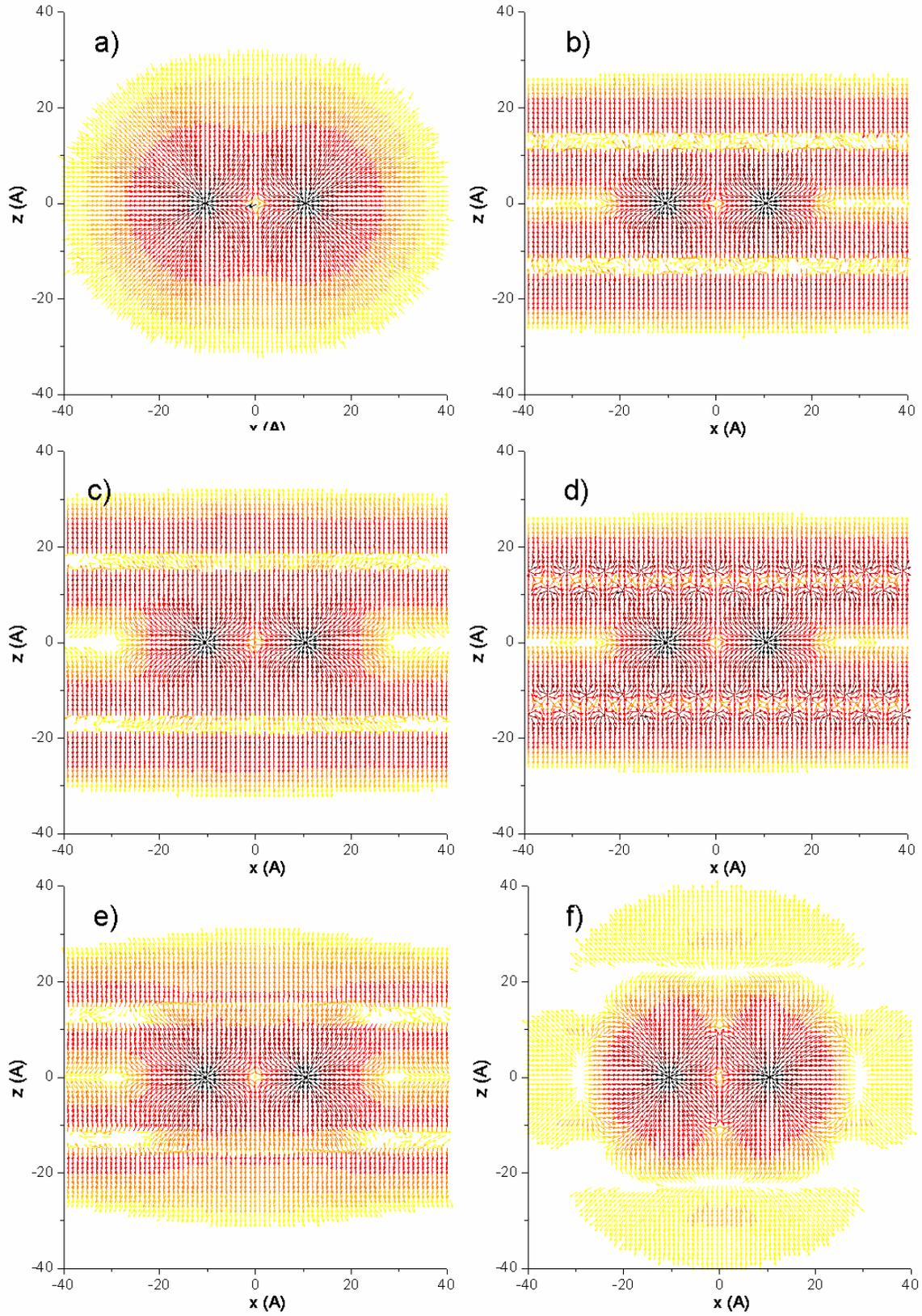}} \caption{\label{Efield} Electric field for the different cases presented along the paper. Arrows indicate the field direction and darker tones points to larger absolute values of it. Tone changes are proportional to $\log(\mid \! \mathbf{E} \! \mid)$ } \end{figure*}

In brief, confinement and self assembling are strongly connected, since attraction between rods and plates depends on a delicate balance between charge and excluded volume of all particles in the system. Additionally, self assembling is linked to several interesting phenomena such as double layer coupling, counterion release, local violations of electroneutrality, and to the appearance of ionic depletion regions. Our belief is that neither of them by itself rules self-assembling, but self-assembling involves all of them acting interdependently, as a whole.

\appendix
\section{Electric field}

We find instructive to show how the electric field looks like for some of the cases here studied. We restrict to show the t=0 configurations, which correspond to the a) plots of figures \ref{rho-sinplan}, \ref{rho-cabmov}, \ref{rho-sep-a}, \ref{rho-cfij}, \ref{rho-perc50}, \ref{rho-perc0}, {\it i.~e.}, to unconfined conditions, confinement by 100$\%$ charged plates at contact with rods, by fully charged plates at $\tau-2R=2a$, by plates with fixed charges, by 50 $\%$ charged plates, and by uncharged plates, all these three last cases for $\tau-2R=0$. The electric field at a given $x$ $y$ position was calculated by simply averaging the electric force per unit of charge acting on charged sites located at the corresponding position. This average is performed for all configurations. For obtaining the electric field for those $x$ and $y$ positions where charged sites cannot access, an extra ghost charged site is included in the calculation. Naturally, this extra ghost charged site is not accounted during the simulation.     

The obtained electric fields are shown in figure \ref{Efield}. Figure \ref{Efield} a) corresponds to the unconfined case. There it is seen that the electric field converges to both rods axes, as expected. The intensity of the electric field decays with the distance to the rods axes, for the region outside the rods, and in a more complicated way for the region in between them. In particular, there is a point of cero electric field at $x=0$ $z=0$, which is noted as lighter tones of the surrounding arrows. 

The electric field is very disturbed by the presence of the confining plates, whether they are charged or not. This is in general seen in figures \ref{Efield} b)-f). Case b), fully charged plates at contact with rods, shows that inside plates the electric field is very week, due to the fact that plate's charge mostly locates at the interface, as commented in section \ref{res}. This fact makes difficult to obtain accurate values for this region, and this is why noise appears in the arrows directions. It should be noted that small positive (repulsive) values for the electric component of the force acting on plates were obtained under these conditions. This agrees with the fact of having a very small electric field inside the plates. In addition, and as expected, the electric field diverges from the plates. This, and the fact of having convergence at the rods axes, produces a very intense local electric field in between rods and plates. There also appears a region of cero electric field at $z=0$, outside the rods. A final thing to note is that the presence of the rods clearly affects the electric field outside the interplate region. 

When plates are a little bit separated from the rods, {\it i.~e.} for $\tau-2R=2a$ (case c)), it is seen that the electric field inside plates increases. This may suggest that larger positive electric force acting on plates should be obtained for this case. However, this force is relatively large but negative (attractive), as mentioned in section \ref{res}. A closer inspection of figure \ref{Efield} c) reveals that arrows diverging from plates toward the midplane are darker than those diverging outside plates. This means that charged sites on the internal plates surfaces suffer larger forces than those at the outer surfaces, which in turn explains the attractive electric force on plates for this case. Additionally, it is seen that these darker arrows are located close to the rods, and lighter arrows are seen diverging from plates for larger distances to the rods. This means that the electric component to the plates pressure is concentrated close to rods, suggesting that plates are supporting shear stress. Hence, non rigid plates such as bilipid membranes may deform to surround rods, as shown experimentally for certain ADN-phospholipid complexes.

Figure \ref{Efield} d) corresponds to the case in which we fixed the charged sites at plates surfaces. These charged sites are clearly distinguished in figure \ref{Efield} d) since they act as divergent sources of electric field. This, in turn, produces the interplate charge density peaks and valeys already shown in figure \ref{rho-cfij}. 

Finally, figures \ref{Efield} e) and f) show the results for plates charged with 50 $\%$ of the full charge, and uncharged plates, respectively. Figure e) shows that the electric field inside the plates points toward the inter rod region, and that arrows that diverge from plates toward the plates midplane are darker than those pointing in the other direction. This fact also well agrees with the fact of obtaining an attractive electric contribution for the pressure acting on plates, as mentioned in section. Again, this case suggests that shear stress is being supported by plates, since the electric field focus just in the region close to rods. On the other hand, figure \ref{Efield} f) shows that the electric field diverges from some space regions which concentrate positive ions. For instance, divergences are seen at $(x=0, z\cong \pm \sqrt{R_a^2-(R+t/2)^2})$, at $(x\cong \pm 25 \hbox{\AA}, z=0)$, and at $z \cong \pm 22 \hbox{\AA}$. These regions show positive charge distributions, as clearly shown by figure \ref{rho-perc0}.


\end{document}